\newcommand{\articletype}{preprint} 
 \documentclass[aps,pre,\articletype,nofootinbib]{revtex4-1}

\usepackage{tikz}
\usepackage{amsmath}
\usepackage{color}
\usepackage{natbib}
\usepackage{cleveref}

 \newcommand{\p}[2]{\frac{\partial{#1}}{\partial{#2}}} 
 
  \newcommand{\pt}[1]{\frac{\partial{#1}}{\partial t}} 
 
 \newcommand{\ve}[1]{\mathbf{#1}}
 
 \newcommand{\f}[2]{\frac{#1}{#2}}
 \newcommand{\half}{\frac{1}{2}}
 \newcommand{\third}{\frac{1}{3}}

\newcommand{\csch}{\text{csch}}
\newcommand{\lap}[1]{\overline{#1}}

\newcommand\numberthis{\addtocounter{equation}{1}\tag{\theequation}}
\begin{document}


\title{Linear motion of multiple superposed viscous fluids} 



\author{Magnus Vartdal}
\email{magnus.vartdal@ffi.no}
\affiliation{Norwegian Defence Research Establishment (FFI), P.O. Box 25, NO-2027 Kjeller, Norway.}

\author{Andreas N. Osnes}
\email{a.n.osnes@its.uio.no}
\affiliation{ Department of Technology Systems, University of Oslo, P.O. Box 70, NO-2007 Kjeller, Norway}



\date{\today}

\begin{abstract}
In this paper the small-amplitude motion of multiple superposed viscous fluids is studied as a linearized initial-value problem. The analysis results in a closed set of equations for the Laplace transformed amplitudes of the interfaces that can be inverted numerically. The derived equations also contain the general normal mode equations, which can be used to determine the asymptotic growth-rates of the systems directly. After derivation, the equations are used to study two different problems involving three fluid layer. The first problem is the effect of initial phase difference on the development of a Rayleigh-Taylor instability and the second is the damping effect of a thin, highly viscous, surface layer. 

\end{abstract}


\maketitle 

\section{Introduction}
The evolution of small-amplitude disturbances on interfaces between viscous fluids is a class of problems that includes the Rayleigh-Taylor (RT) \cite{Taylor1950,Sharp1984,Kull1991} and Richtmyer-Meshkov type instabilities \cite{Richtmyer1960,Meshkov1969}, as well as damped oscillatory waves \cite{Harrison1908,Lamb1932}.
This study investigates the motion of interface perturbations in the presence of multiple interfaces. The systems considered are subject to continuous acceleration, and thus, depending on the configuration, each interface can be RT unstable or stable and damped. 

The RT instability occurs when a dense fluid is accelerated into a lighter fluid. It plays a dynamically important role in a vast number of natural phenomena ranging in size from cellular level bioconvection \cite{Plesset1974} to nebula formation \cite{Ribeyre2004}. It also occurs as a limiting factor in inertial confinement fusion \cite{Freeman1977,Wouchuk1995,Atzeni2004}. In spherical detonations, the RT instability occurs together with the Richtmyer-Meshkov instability, which is its impulsive analogue. These two instabilities are the driving mechanisms by which the detonation products are mixed with ambient air \cite{Frost2005}. Explosives with poor oxygen balance release more energy as a result of this mixing. Further examples of RT applications can be found in the extensive review of \citet{Zhou2017c,Zhou2017d}.

The opposite case, where a light fluid is accelerated into a denser fluid, is stable and typically results in damped oscillatory wave motion \cite{Harrison1908,Lamb1932,Prosperetti1981}. These waves display a remarkable range of scales, from large tidal waves and tsunamis down to capillary waves driven by surface tension. The damping rate in some of these systems is known to be significantly affected by the presence of surface films and thin surface layers of another fluid \cite{Miles1967,Buckmaster1973, Jenkins1997,Jenkins1997a}. The enhanced damping of such surface layers reduces radar backscatter, which makes it possible to detect oil spills remotely \cite{Alpers1988,Alpers2017}. Furthermore, viscous fluid surface layers have been successfully used to model the damping of ocean waves caused by the presence of ice \cite{Weber1987}.

Traditionally, the evolution of interface perturbations in the linear regime has been investigated by means of normal-mode analysis \cite{Harrison1908,Lamb1932,Chandrasekhar2013}, which is well suited for studying the asymptotic behavior of such systems. Normal-mode analysis can, however, be impractical to use for capturing initial transients. Laplace transform based techniques are better suited for this purpose, since they naturally account for the growth of all modes. This is particularly true for stable configurations where such transients are known to persist for a significant amount of time. For the single interface case, such initial-value problems have been investigated using Laplace transform based techniques \cite{Carrier1959,Prosperetti1976,Menikoff1978,Prosperetti1980,Prosperetti1981,Berger1988,Denner2016}. These problems are commonly used as verification cases for multiphase flow codes \cite{Herrmann2008}.

The presence of nearby interfaces, or a finite fluid layer thickness, can have a substantial effect on the evolution of disturbances, and such multilayer configurations have received considerable theoretical attention \cite{Mikaelian1982,Mikaelian1982a,Mikaelian1990a,Mikaelian1990b,Yang1993,Mikaelian1996,Jenkins1997,Goncharov2000,Mikaelian2005,Piriz2018}. These studies cover both inviscid and viscous cases, but no exact linear theory for an arbitrary number of viscous fluids is available.   

Experimental investigations of unstable multi-layer configurations are challenging due to the difficulty of setting up such systems. To the authors knowledge the only two studies that have done this are the study of \cite{Jacobs2005} and the recent study of \cite{Adkins2017}. Only the latter study could control the initial perturbations, enabling a comparison with the inviscid multi-layer theory of \citet{Mikaelian1990b}. The experimental results were later compared to a viscous three layer solution (limited to two viscous fluids and one free boundary) \cite{Piriz2018}. It was demonstrated that the growth-rate in the experiments were significantly lower than predicted by viscous theory. A possible explanation for the discrepancy is the limited depth of the cell used to conduct the experiments (around $1/4$th the wave length for the shortest wave length considered). A rough estimate of the importance of the viscous effects associated with the cell thickness reveals that they are, at best, of the same order as those included in the theory. Due to the lack of experimental data, and the approximations made by previous theoretical studies, the knowledge about the properties of unstable viscous multi-layer systems is currently limited. 

In this paper, we consider the small-amplitude motion of an arbitrary number of superposed viscous fluids as a linearized initial-value problem. To our knowledge, this is the first study to approach the multi-layer problem in this fashion. The present work is an extension of the single interface analysis of \citet{Prosperetti1981}. The procedure results in a closed set of equations, involving only the Laplace transformed amplitudes of the interfaces, which can be inverted numerically. As in \cite{Prosperetti1981}, we assume, for simplicity, that no vorticity is present initially.
The derived equations also contain the general normal mode equations for an arbitrary number of viscous fluids. As far as the authors know this relation is also novel. 
 
After deriving the equations we use them to study two different three-layer problems. The first problem is the effect of initial phase difference on the development of an RT instability, and the second is the damping effect of a thin highly-viscous surface layer.

\section{Problem formulation and decomposition} 
\begin{figure} 
  \ifnum\pdfstrcmp{\articletype}{preprint}=0 
    \includegraphics{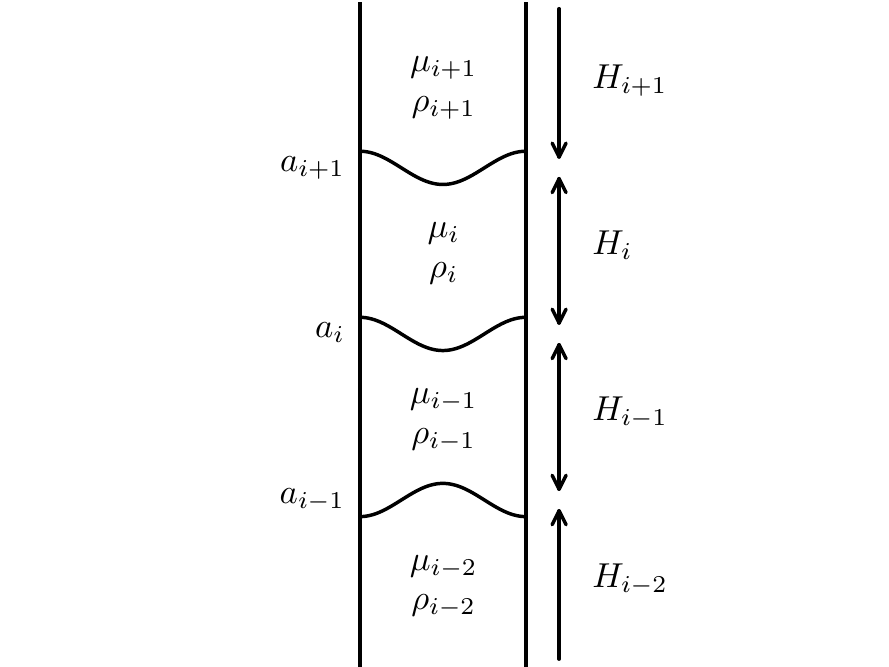}
    \else 
    \includegraphics{figure1.eps}  
    \fi
 \caption{Schematic illustration of the multi-layered initial-value problem. $a_i$ is the amplitude of the disturbance on interface $i$. $H_i$, $\rho_i$, and $\mu_i$, are the thickness, density, and dynamic viscosity of fluid $i$, respectively.}\label{Fig:Config}
\end{figure}

Consider a configuration of N+1 superposed incompressible viscous fluids separated by N interfaces, where interface $i$ separates fluid $i-1$ and $i$, as depicted in Figure \ref{Fig:Config}. Fluid layer $i$ has constant thickness, density, and dynamic viscosity denoted by $H_i$, $\rho_i$, and $\mu_i$, respectively. The coordinate system is oriented such that the equilibrium position of each interface is given by $y_i=constant$, and gravity, denoted by $g$, acts opposite the $y$-axis. Initially, each interface is perturbed around its equilibrium position in an arbitrary manner, but since we restrict our analysis to the linear regime, these perturbations can be decomposed into separate modes by means of a Fourier transform. With this transformation, the equations describing the interfaces can be expressed as  
\begin{equation} \label{surface-form}
 \eta_i(x,z,t)=a_i(t)f(x,z)+y_i,
\end{equation}
where $a_i$ is the amplitude of the disturbance, and $f$ satisfies the Helmholtz equation 
\begin{equation} \label{Helmholtz}
\left(\f{\partial^2}{\partial x^2}+\f{\partial^2}{\partial z^2}+k^2\right)f=0, 
\end{equation}
where $k=(k_x^2+k_z^2)^\half$ is the wavenumber of the disturbance. In the remainder of the paper, subscripts are dropped for convenience when no confusion can arise.

The motion of each fluid is governed by the linearized Navier-Stokes equations,
\begin{equation} \label{Lin-N-S}
 \pt{\ve{u}}=-\f{1}{\rho}\nabla p +\nu\nabla^2\ve{u}+\ve{g}.
\end{equation}
Here, $\ve{u}$ is the velocity, $p$ is the pressure, and $\nu=\mu / \rho$ is the kinematic viscosity of the fluid. At the interfaces, the linearized kinematic and dynamic boundary conditions, including the effects of surface tension, are enforced. In the general case, the linearization requires that the amplitude at each interface is small compared to both the wavelength $\lambda=2 \pi/k$ and the thickness of the surrounding layers, i.e, $a_i<<\lambda, H_i,H_{i-1}$.\footnote{There are exceptions where the conditions of linearity are less strict. For instance, a thin film on top of a thick fluid layer with wave amplitudes larger than the film thickness ($a>>H$) can be treated linearly if the waves are long \cite{Jenkins1997}.}
 
To solve eqns \eqref{surface-form}, \eqref{Helmholtz}, and \eqref{Lin-N-S}, the decomposition procedure found in \cite{Prosperetti1981} is used. Some of the details of the procedure is repeated here for the readers convenience. First, the pressure and volume force terms are eliminated by applying the curl operator to the linearized Navier-Stokes equation. This results in 
\begin{equation} \label{Lin-vort}
 \pt{\ve{\omega}}=\nu\nabla^2\ve{\omega},
\end{equation}
where $\ve{\omega}$ is the vorticity of the fluid. Since the vorticity is divergence free by definition, it can be represented by a vector potential of the form $\ve{\omega}=\nabla \times(\ve{A}+\nabla\times \ve{B})$. This decomposition is well suited for the present problem, since it has been demonstrated that $\ve{A}$ and $\ve{B}$ can be reduced to single component form by means of a gauge transformation \cite{Cortelezzi1981}. The resulting vectors can be expressed as
\begin{align*} 
\ve{A}=[0,\,\, \Omega(y,t)f(x,z),\,\,0],\\ 
\ve{B}=[0,\,\, G(y,t)f(x,z),\,\,0]. \numberthis \label{AandB-comp}
\end{align*}
Introducing \eqref{AandB-comp} into \eqref{Lin-vort}, and employing the Helmholtz equation \eqref{Helmholtz}, we find the evolution equation for $\Omega$   
\begin{equation} \label{Omega_eq}
 \left[\p{}{t}-\nu(\p{^2}{y^2}-k^2)\right] \Omega=0.
\end{equation}
The evolution equation for $G$ is on the same form.

While $\ve{A}$ and $\ve{B}$ are sufficient for a complete description of the vorticity, an additional scalar potential, $\phi$, is required to represent the velocity. With this addition, the velocity can be expressed as
\begin{equation} \label{vel-decomp}
\ve{u}=\ve{A}+\nabla \times \ve{B}-\nabla \phi.
\end{equation}
Introducing \eqref{vel-decomp} into the vertical component of the linearized Navier-Stokes equation \eqref{Lin-N-S}, and employing \eqref{Omega_eq} and \eqref{Helmholtz}, results in a Bernoulli type equation for the pressure
\begin{equation} \label{Bernoulli}
 p=-\rho g y +\rho\p{\phi}{t}-\mu\p{\Omega}{y}f+C,
\end{equation}
where $C$ is a constant. Further specification of the pressure requires knowledge about the scalar potential. The required equation for $\phi$ is obtained from the incompressibility constraint 
\begin{equation} \label{Poisson}
\nabla^2 \phi=\nabla \cdot \ve{A}=\p{\Omega}{y}f.
\end{equation}
Introducing the decomposition $\phi=\Phi(y,t)f(x,z)$ one finds
\begin{equation} \label{Phi_eq}
  \left(\p{^2}{y^2}-k^2\right) \Phi=\p{\Omega}{y}.
\end{equation}
The general solution to \eqref{Phi_eq} can be found using Lagrange's method of variation of parameters. 

Up until this point, the analysis is identical to that presented by Prosperetti \cite{Prosperetti1981} for the single interface case. The introduction of more interfaces does, however, alter the form of the scalar potential, as the kinematic boundary condition  
\begin{equation}
 \p{\Phi}{y}=\Omega -\dot{a},
\end{equation}
must be enforced on each interface. Here, $\dot{a}$ denotes the time derivative of the amplitude. The complete expression for the scalar potential is rather complicated, but for the remaining analysis only the expression for the potential at the interface locations are needed. At the interface locations the expression simplifies to
\begin{align}
\Phi_i (y_i)=\coth(kH_i) k^{-1}\dot{a}_i  - \csch(kH_i) k^{-1}\dot{a}_{i+1}\nonumber\\ 
-\int_{y_i}^{y_{i+1}}\Omega_i \frac{\sinh (k(y_{i+1}-y))}{\sinh (kH_i)}dy \nonumber \\  
\Phi_i (y_{i+1})=\csch(kH_i) k^{-1}\dot{a}_i  - \coth (kH_i) k^{-1}\dot{a}_{i+1}\nonumber\\ 
+\int_{y_i}^{y_{i+1}}\Omega_i \frac{\sinh(k(y-y_i))}{\sinh (kH_i)}dy \label{phiy1},
\end{align}
where the subscript on $\Phi$ denotes which fluid layer the potential is defined in. From the requirement of continuity of tangential velocity at the interfaces it follows that
\begin{equation} \label{Phiy}
 \Phi_{i}(y_i,t)=\Phi_{i-1}(y_i,t), 
\end{equation}
with an identical relation for $G_i$. Substituting \eqref{phiy1} into \eqref{Phiy} yields the condition 
\ifnum\pdfstrcmp{\articletype}{preprint}=0
\begin{multline}
  \Big(\coth(kH_{i-1})+\coth(kH_i)\Big)\dot{a}_i 
- \csch(kH_{i-1})\dot{a}_{i-1}-\csch(kH_i)\dot{a}_{i+1} \\
= k\left(\int_{y_{i-1}}^{y_{i}}\Omega_{i-1} \frac{\sinh (k(y-y_{i-1}))}{\sinh (kH_{i-1})}dy+\int_{y_i}^{y_{i+1}}\Omega_i \frac{\sinh (k(y_{i+1}-y))}{\sinh (kH_i)}dy\right) \label{stress_omega_eq},
\end{multline}
\else
\begin{multline}
  \Big(\coth(kH_{i-1})+\coth(kH_i)\Big)\dot{a}_i- \csch(kH_{i-1})\dot{a}_{i-1} \\
-\csch(kH_i)\dot{a}_{i+1}= k\Big(\int_{y_{i-1}}^{y_{i}}\Omega_{i-1} \frac{\sinh (k(y-y_{i-1}))}{\sinh (kH_{i-1})}dy \\
+\int_{y_i}^{y_{i+1}}\Omega_i \frac{\sinh (k(y_{i+1}-y))}{\sinh (kH_i)}dy\Big) \label{stress_omega_eq},
\end{multline}
\fi
which couples the velocities of adjacent interfaces. The continuity of tangential stresses yield the same equations as those reported in \cite{Prosperetti1981}, i.e,
\begin{equation} \label{Omega_tan_eq}
 \mu_{i}\Omega_i(y_i,t)-\mu_{i-1}\Omega_{i-1}(y_{i},t)=2(\mu_{i}-\mu_{i-1})\dot{a}_{i},
\end{equation}
\begin{equation}
 \p{}{y}\Big(\mu_i G_i(y_i,t)-\mu_{i-1} G_{i-1}(y_i,t)\Big)=0.
\end{equation}
The continuity of normal stress can be simplified to 
\begin{equation}
-p_{i}+p_{i-1}-2k^2(\mu_{i}\Phi_i+\mu_{i-1}\Phi_{i-1})=ak^2\zeta,
\end{equation}
where $\zeta$ is the surface tension coefficient. Next, the pressure is eliminated using \eqref{Bernoulli} followed by the elimination of the scalar potential using \eqref{phiy1}. This yields evolution equations for the amplitudes which only depend on the $\Omega_i$-fields and the amplitudes themselves,
\ifnum\pdfstrcmp{\articletype}{preprint}=0
\begin{multline}
m_i\ddot{a}_i+c_i\dot{a}_i+d_i a_i=\csch (kH_{i-1})(\rho_{i-1}\ddot{a}_{i-1}+2 \mu_{i-1}k^2 \dot{a}_{i-1})+\csch (kH_{i})(\rho_{i}\ddot{a}_{i+1}+2 \mu_{i}k^2 \dot{a}_{i+1})\\
-k^2 \mu_{i-1}\Big(\Omega_{i-1}(y_i,t)\coth(kH_{i-1})-2k\int_{y_{i-1}}^{y_{i}}\Omega_{i-1} \frac{\sinh (k(y-y_{i-1}))}{\sinh (kH_{i-1})}dy-\Omega_{i-1}(y_{i-1},t)\csch (kH_{i-1})\Big)\\
-k^2 \mu_{i}\Big(\Omega_{i}(y_i,t)\coth(kH_{i})-2k\int_{y_{i}}^{y_{i+1}}\Omega_{i}\frac{\sinh (k(y_{i+1}-y))}{\sinh (kH_{i})}dy-\Omega_i(y_{i+1},t)\csch (kH_{i})\Big).\\
\label{normeq}
\end{multline}
\else
\begin{widetext}
\begin{multline}
m_i\ddot{a}_i+c_i\dot{a}_i+d_i a_i=\csch (kH_{i-1})(\rho_{i-1}\ddot{a}_{i-1}+2 \mu_{i-1}k^2 \dot{a}_{i-1})+\csch (kH_{i})(\rho_{i}\ddot{a}_{i+1}+2 \mu_{i}k^2 \dot{a}_{i+1})\\
-k^2 \mu_{i-1}\Big(\Omega_{i-1}(y_i,t)\coth(kH_{i-1})-2k\int_{y_{i-1}}^{y_{i}}\Omega_{i-1} \frac{\sinh (k(y-y_{i-1}))}{\sinh (kH_{i-1})}dy-\Omega_{i-1}(y_{i-1},t)\csch (kH_{i-1})\Big)\\
-k^2 \mu_{i}\Big(\Omega_{i}(y_i,t)\coth(kH_{i})-2k\int_{y_{i}}^{y_{i+1}}\Omega_{i}\frac{\sinh (k(y_{i+1}-y))}{\sinh (kH_{i})}dy-\Omega_i(y_{i+1},t)\csch (kH_{i})\Big).\\
\label{normeq}
\end{multline}
\end{widetext}
\fi
Here, $m_i=\rho_{i-1}\coth(kH_{i-1})+\rho_i\coth(kH_i)$, $c_i=2k^2(\mu_{i-1}\coth(kH_{i-1})+\mu_i\coth(kH_i))$, $d_i=(\rho_{i-1}+\rho_i)\omega_i^2$, and $\omega_i$ can be recognized as the inviscid natural frequency for an interface separating two infinite fluid layers
\begin{equation} \label{invicidOmega}
\omega_i^2=\f{\rho_{i-1}-\rho_{i}}{\rho_{i-1}+\rho_{i}}gk+\f{\zeta}{\rho_{i-1}+\rho_{i}}k^3.
\end{equation}
It is readily seen that, in the limit of infinite layer thickness ($kH\rightarrow\infty$) equations \eqref{stress_omega_eq} and \eqref{normeq} simplify to the corresponding equations for the single interface case, and we obtain the same set of equations as Prosperetti \cite{Prosperetti1981}.

It should be noted that, just like the single interface case, the evolution of the $G$-component of the vorticity is decoupled from that of the amplitudes and $\Omega$. Since we have limited our study to the vanishing initial vorticity case, $G$ does not enter in the evolution equations in any way. Therefore, it is not considered further.

\section{Laplace transformed equations of motion}
Unlike the single interface case, no closed form solution of the above time domain equations have been found. It is, however, possible to obtain a closed set of equations involving only the amplitudes after Laplace transformation of equations \eqref{Omega_eq}, \eqref{stress_omega_eq}, \eqref{Omega_tan_eq}, and \eqref{normeq}. 

Let an overbar indicate a Laplace transformed quantity, $s$ denote the frequency parameter and define  
\begin{equation}\label{lambda_def}
\lambda_{i}=(k^2+s/\nu_{i})^\half.
\end{equation}
Using this variable, the general Laplace transformed solution of equation \eqref{Omega_eq} can be expressed
\begin{equation}\label{Omega_lap}
\lap{\Omega}_i=A_i\frac{\sinh(\lambda_i(y-y_i))}{\sinh(\lambda_i H_i)}+B_i\frac{\sinh(\lambda_i(y_{i+1}-y))}{\sinh(\lambda_i H_i)}.
\end{equation}
The tangential conditions on the interface, \eqref{stress_omega_eq} and \eqref{Omega_tan_eq}, yield a set of equations for determining the coefficients $A_i$ and $B_i$ in terms of $\lap{\dot{a}}$:
\ifnum\pdfstrcmp{\articletype}{preprint}=0
\begin{multline}
\gamma_{i-1}A_{i-1}+\delta_{i-1}B_{i-1}+\delta_{i}A_{i}+\gamma_{i}B_{i}=\\
k^{-1}\Big[\Big(\coth(kH_{i-1})+\coth(kH_i)\Big)\lap{\dot{a}}_i-\csch(kH_{i-1})\lap{\dot{a}}_{i-1}-\csch(kH_i)\lap{\dot{a}}_{i+1}\Big],\\ \label{lap_omega_stress}
\end{multline}
\else
\begin{multline}
\gamma_{i-1}A_{i-1}+\delta_{i-1}B_{i-1}+\delta_{i}A_{i}+\gamma_{i}B_{i}=k^{-1}\Big[\Big(\coth(kH_{i-1})\\
+\coth(kH_i)\Big)\lap{\dot{a}}_i-\csch(kH_{i-1})\lap{\dot{a}}_{i-1}-\csch(kH_i)\lap{\dot{a}}_{i+1}\Big],\\\label{lap_omega_stress}
\end{multline}
\fi
\begin{equation}\label{lap_omega_tan}
\mu_i B_i-\mu_{i-1}A_{i-1}=2(\mu_i-\mu_{i-1})\lap{\dot{a}}_i,
\end{equation}
where $\gamma$ and $\delta$ are given by 
\begin{align} 
 \gamma_i=\frac{1}{\lambda_i^2-k^2}(\lambda_i \coth(\lambda_iH_i)-k\coth(k H_i)),\nonumber\\
 \delta_i=\frac{1}{\lambda_i^2-k^2}(k \csch(kH_i)-\lambda_i\csch(\lambda_i H_i)).\label{gamma-delta}
\end{align}
The solution to the above set of equations can be substituted into the Laplace transformed normal stress equation \eqref{normeq}, resulting in
\ifnum\pdfstrcmp{\articletype}{preprint}=0
\begin{multline}
 (m_is^2+c_is+d_i)\lap{a}_i-\csch(kH_{i-1})(\rho_{i-1}s^2+2\mu_{i-1}k^2s)\lap{a}_{i-1}-\csch(kH_i)(\rho_is^2+2\mu_{i}k^2s)\lap{a}_{i+1}\\
 +\mu_{i-1}k^2(\beta_{i-1}A_{i-1}+\alpha_{i-1}B_{i-1})+\mu_{i}k^2(\alpha_{i}A_{i}+\beta_{i}B_{i})=\frac{1}{s}(m_is^2+c_is+d)a_i^0-\frac{d}{s}a_i^0+m_iu_i^0\\
 -\csch(kH_{i-1})((\rho_{i-1}s+2\mu_{i-1}k^2)a_{i-1}^0+\rho_{i-1}u_{i-1}^0)-\csch(kH_{i})((\rho_{i}s+2\mu_{i}k^2)a_{i+1}^0+\rho_{i}u_{i+1}^0),\label{governing}
\end{multline}
\else
\begin{widetext}
\begin{multline}
 (m_is^2+c_is+d_i)\lap{a}_i-\csch(kH_{i-1})(\rho_{i-1}s^2+2\mu_{i-1}k^2s)\lap{a}_{i-1}-\csch(kH_i)(\rho_is^2+2\mu_{i}k^2s)\lap{a}_{i+1}\\
 +\mu_{i-1}k^2(\beta_{i-1}A_{i-1}+\alpha_{i-1}B_{i-1})+\mu_{i}k^2(\alpha_{i}A_{i}+\beta_{i}B_{i})=\frac{1}{s}(m_is^2+c_is+d)a_i^0-\frac{d}{s}a_i^0+m_iu_i^0\\
 -\csch(kH_{i-1})((\rho_{i-1}s+2\mu_{i-1}k^2)a_{i-1}^0+\rho_{i-1}u_{i-1}^0)-\csch(kH_{i})((\rho_{i}s+2\mu_{i}k^2)a_{i+1}^0+\rho_{i}u_{i+1}^0),\label{governing}
\end{multline}
\end{widetext}
\fi
where $a_i^0$ and $u_i^0$ are the initial amplitude and velocity of interface $i$, respectively and $\alpha_i$ and $\beta_i$ are given by 
\begin{align} \label{alpha-beta}
 \alpha_i=\frac{1}{\lambda_i^2-k^2}(2k\lambda_i \csch(\lambda_iH_i)-(\lambda_i^2+k^2)\csch(k H_i)),\nonumber\\
 \beta_i=\frac{1}{\lambda_i^2-k^2}((\lambda_i^2+k^2)\coth(kH_i)-2k\lambda_i\coth(\lambda_i H_i)). 
\end{align}
Equations \eqref{governing}, and \eqref{alpha-beta} form a closed set of equations for the Laplace transform of the amplitudes, which can be inverted to find the evolution of the interfaces in time. The final step cannot be handled analytically, and a numerical inverse Laplace transform algorithm is required. For long time integration this algorithm can be very sensitive to numerical precision issues. In these cases we have employed arbitrary precision versions of the Euler and Talbot algorithms. For a description of the algorithms the reader can consult \cite{Abate2006}.

In the limit of infinite layer thickness ($kH\rightarrow \infty $), the above equations simplify considerably. Take, for instance, the case of a bottom layer of infinite depth and let $a_1$ represent the lowest interface amplitude. The equations for $a_1$ can be simplified using the following relations: 
$\coth(kH_0)\rightarrow 1$, $\csch(kH_0)\rightarrow 0$, $ B_0\rightarrow0$, $\alpha_0\rightarrow 0$, $\beta_0\rightarrow (\lambda_0-k)/(\lambda_0+k)$,
$\delta_0\rightarrow 0$ and $\gamma_0\rightarrow 1/(\lambda_0+k)$.
With these simplifications all references to $a_0$ disappear and the system is closed. For a top layer of infinite extent similar relations apply, with the exception that it is the $A$ coefficient and not the $B$ that disappears in the top layer. 

For the case of finite depth above a fixed wall, the tangential stress condition at the wall \eqref{lap_omega_tan} can no longer be used. The tangential velocity conditions reduces to 
\begin{equation}
\delta_0 A_0 + \gamma_0 B_0=-k^{-1}\csch (kH_0) \lap{\dot{a}}_1,
\end{equation}
as a consequence of $\Phi (y_0)=0$ at the wall.

Another interesting limit is the interaction of a highly viscous fluid with other fluids of very low viscosity. In the limit of zero viscosity, the above equations become ill-defined because $\lambda \rightarrow \infty$. However, if the initial vorticity is zero, Kelvin's circulation theorem ensures that vorticity remains identically zero. This implies that $A$ and $B$ remain 0 for the inviscid fluid. The continuity of tangential velocity, \eqref{lap_omega_stress}, should not be applied at these interfaces.
    
It should also be pointed out that if all terms involving initial-values are removed from \eqref{governing}, \eqref{lap_omega_stress} and \eqref{lap_omega_tan}.  The remaining equation system represents the normal-mode equations for the given initial-value problem. Within this interpretation $s$ represents the growth rate of the normal mode, and $\lap{a}_i$ is the associated eigenvector. The above equation is thus also useful for evaluating the asymptotic behavior of the system directly. In the limiting case of infinite fluid thickness the expression for the growth-rate reduces to that of the normal-modes found in \cite{Bellman1954}. Furthermore, we note that the initial behavior of the system, also known as the irrotational approximation, is found by setting A and B equal to zero in all fluid layers. 

\section{Results}

\subsection{Initial phase effects}
One of the topics that motivated this work was the effect of nearby interfaces on the evolution of a Rayleigh-Taylor instability and in particular what a difference in initial phase between the interfaces could result in.

For multi-layer cases, the number of parameters needed to describe a given configuration quickly becomes exceedingly large. We have thus chosen to restrict our study to the case of a single finite layer trapped between two semi-infinite fluids, but even for this very limited case 12 non-dimensional parameters are needed to classify the problem. We therefore further restrict our cases by neglecting surface tension effects, assuming zero initial velocity, equal Atwood numbers for the two interfaces, and equal kinematic viscosities for all fluids. These assumptions reduce the number of non-dimensional parameters to $4$. We chose the following parameters: the amplitude ratio, $a_r=a^0_2/a_1^0$, non-dimensional layer thickness $h=kH$, Atwood number $A=(\rho_2-\rho_1)/(\rho_2+\rho_1)$, and the viscosity parameter $\epsilon=\nu k^2/|\omega_1^2|^\half$ used in \cite{Prosperetti1981}. Of these parameters the last two characterize the material properties, and for these we have used a fixed set of three values each. The chosen values, $A \in (0.1, 0.5, 0.9)$ and $\epsilon \in (1, 0.1, 10^{-3})$,  represent the low, medium, and high end of each parameter space. For all cases the viscous time scale from \cite{Menikoff1977} ($T=(\nu/A^2g^2)^\third$) is used to construct a non-dimensionalized time, $\tau=t/T$.

\begin{figure} 
  \ifnum\pdfstrcmp{\articletype}{preprint}=0 
    \includegraphics[width=0.75\linewidth]{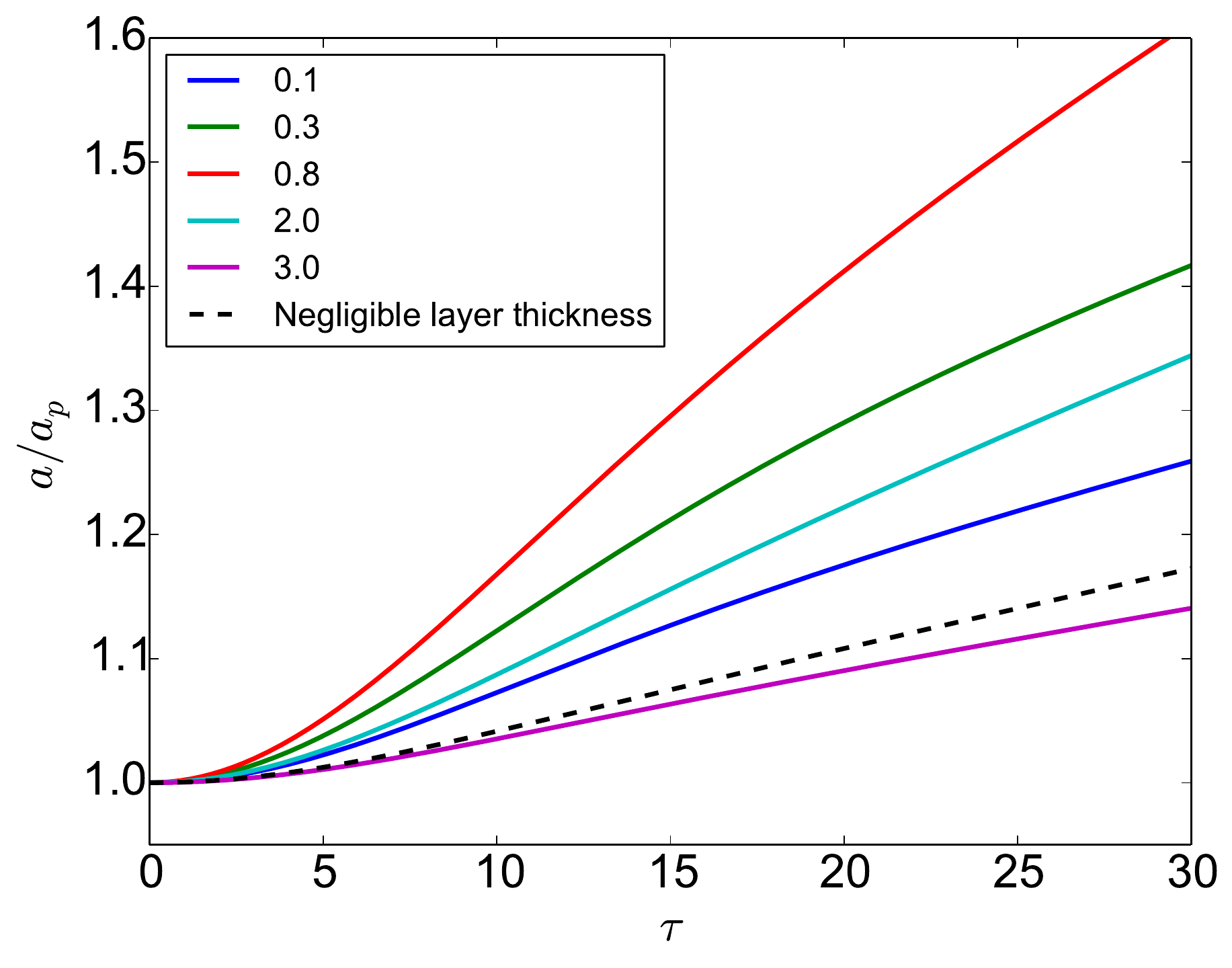}  
    \else 
    \includegraphics[width=\linewidth]{figure2.pdf}  
    \fi
 \caption{The amplitudes of the disturbances on the lower interface for the case $A=0.9$, $\epsilon=10^{-3}$, and $a_r=1$, normalized by the infinite layer thickness solution $a_p$. The legend denotes the non-dimensional layer thickness $h$.}\label{Fig:unstable1}
\end{figure}

First, we consider the case when $a_r=1$, i.e. when the interfaces are initially in phase. At first glance, one might think that the resulting time histories, for both interfaces, would be bounded by the solutions for the asymptotic cases of infinite layer height and negligible layer height, which have analytic solutions \cite{Prosperetti1981}. This is indeed the case for the upper interface between the two densest fluids. However, as seen in Figure \ref{Fig:unstable1}, for the case $A=0.9$ and $\epsilon=10^{-3}$, the amplitude of the lower interface initially grows faster than the asymptotic case of negligible fluid height when $h<3$, with a maximum at $h=0.8$ (for the range of $h$ shown in the figure). We observe the same non-monotonic behavior for all the cases with $A=0.9$ regardless of which viscosity parameter is used (results not shown).

\begin{figure} 
  \ifnum\pdfstrcmp{\articletype}{preprint}=0 
    \includegraphics[width=0.75\linewidth]{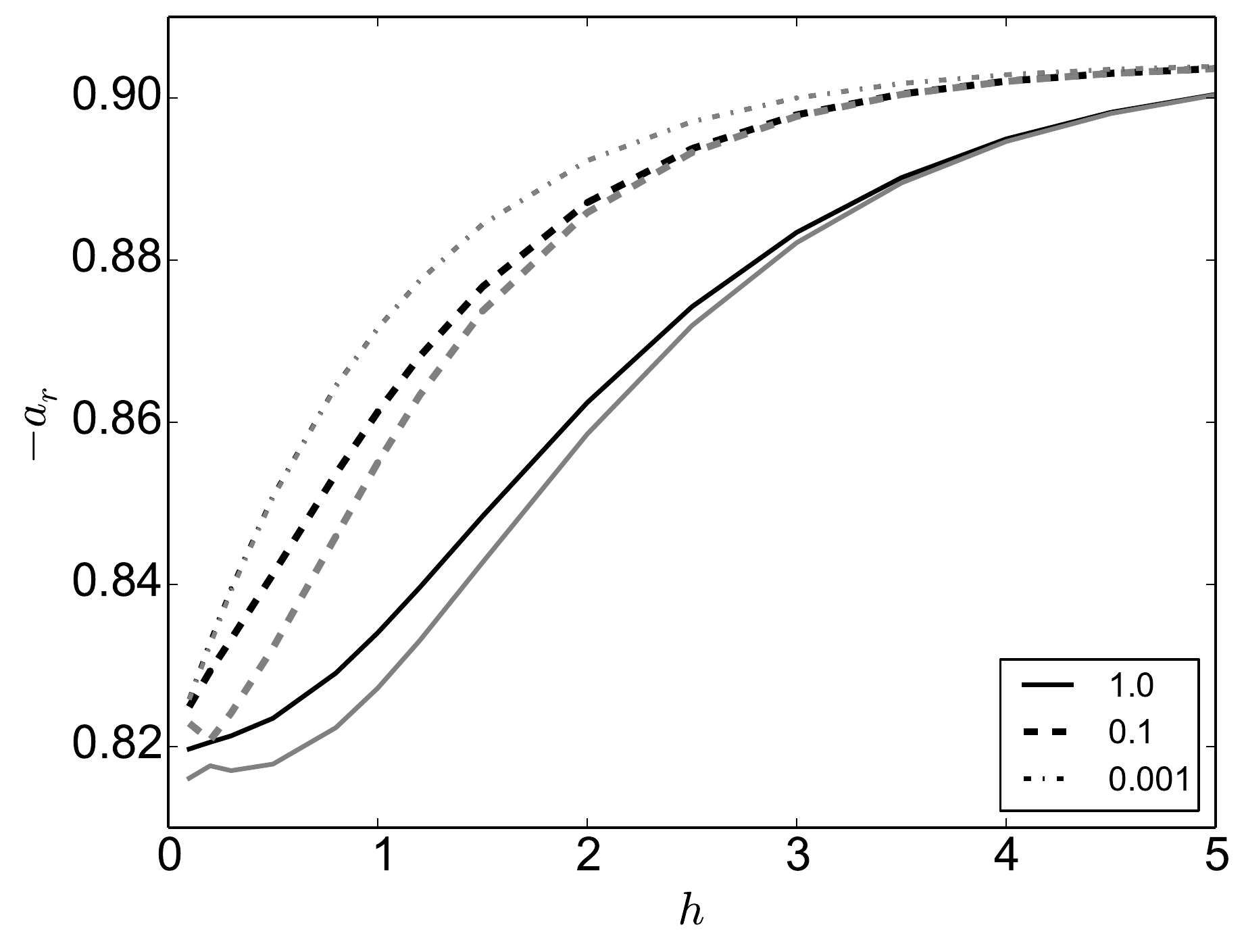}  
    \else 
    \includegraphics[width=\linewidth]{figure3.pdf}  
    \fi
 \caption{Critical amplitude ratio as a function of non-dimensional layer thickness for $A=0.1$. Initial amplitude ratio (black) and normal mode ratio (gray). The legend denotes the value of the viscosity parameter $\epsilon$.}\label{Fig:ArPlot01}
\end{figure}

\begin{figure} 
  \ifnum\pdfstrcmp{\articletype}{preprint}=0 
    \includegraphics[width=0.75\linewidth]{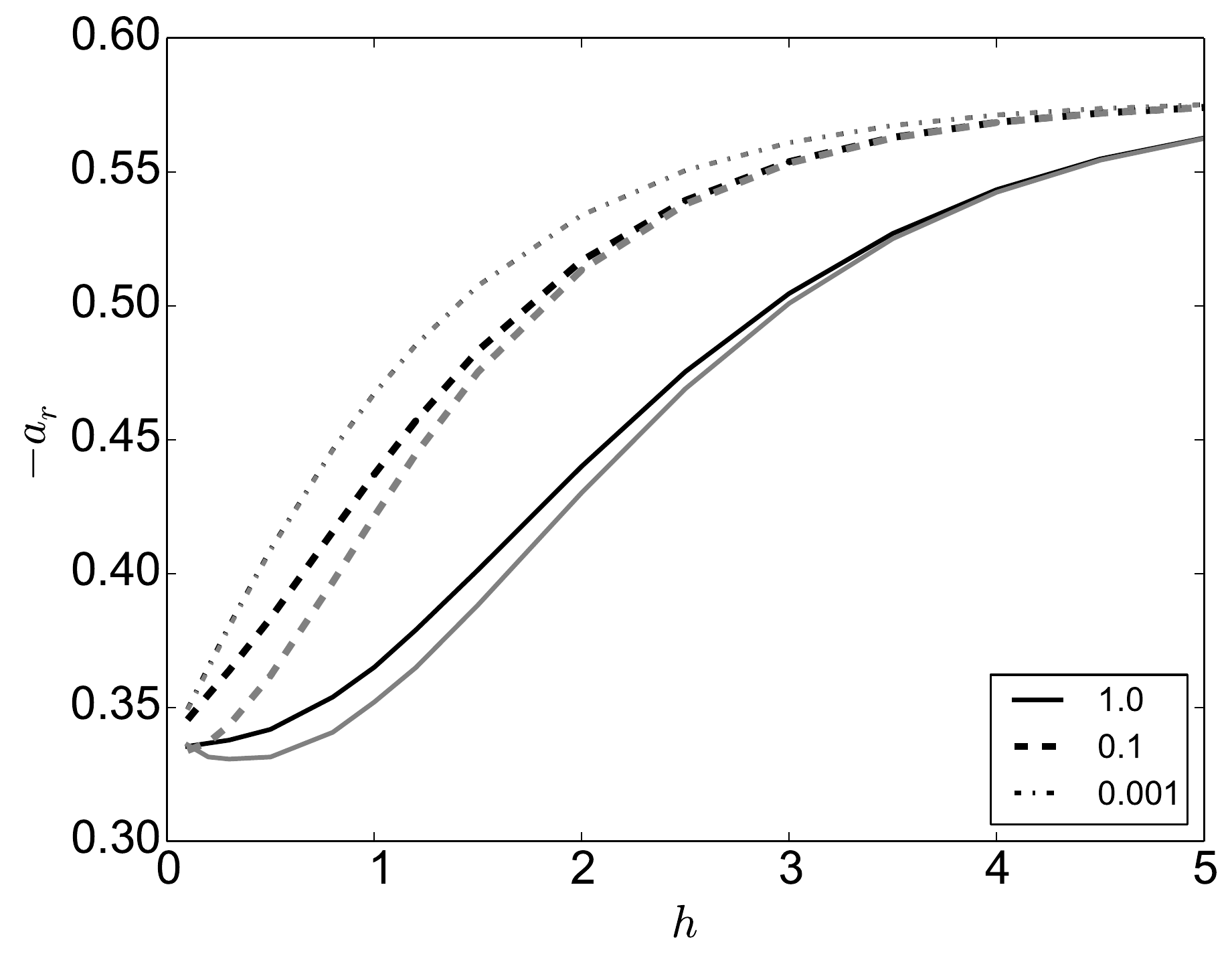}  
    \else 
    \includegraphics[width=\linewidth]{figure4.pdf}  
    \fi
 \caption{Critical amplitude ratio as a function of non-dimensional layer thickness for $A=0.5$. Initial amplitude ratio (black) and normal mode ratio (gray). The legend denotes the value of the viscosity parameter $\epsilon$.}\label{Fig:ArPlot05}
\end{figure}

\begin{figure} 
  \ifnum\pdfstrcmp{\articletype}{preprint}=0 
    \includegraphics[width=0.75\linewidth]{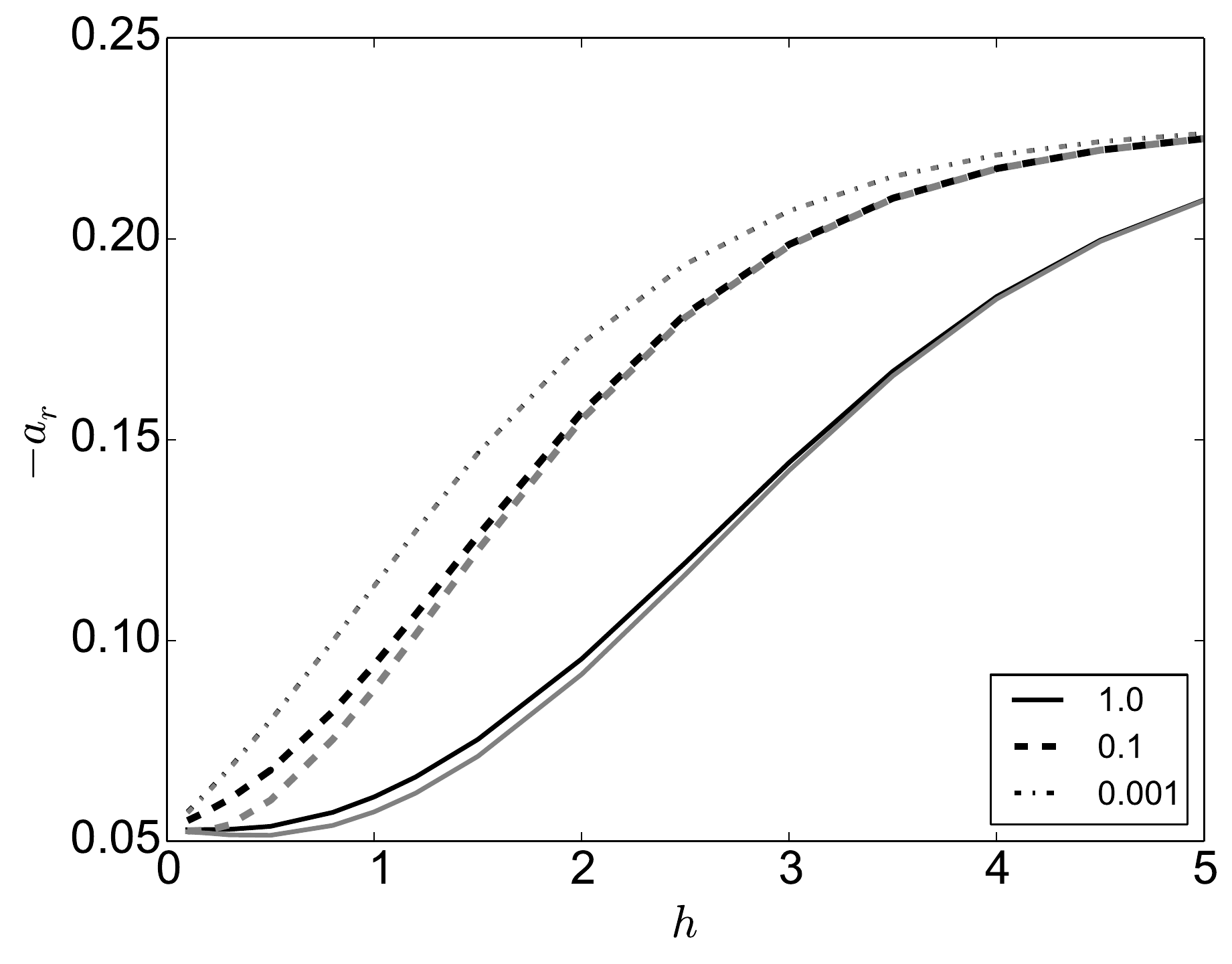}  
    \else 
    \includegraphics[width=\linewidth]{figure5.pdf}  
    \fi
 \caption{Critical amplitude ratio as a function of non-dimensional layer thickness for $A=0.9$. Initial amplitude ratio (black) and normal mode ratio (gray). The legend denotes the value of the viscosity parameter $\epsilon$.}\label{Fig:ArPlot09}
\end{figure}

For the case $a_r=-1$, the resulting motion is first for the interfaces to move in opposite directions. Eventually, however, the motion becomes dominated by the interface between the heavier fluids and both move together at the same asymptotic growth-rate. The change in direction for the lower interface, which does not happen when $a_r=1$, indicates that there should exist a minimal amplitude ratio for which a reversal of motion of the lower interface does not occur. This ratio is where a minimal growth-rate of the disturbances is realized, for a given set of $A$, $\epsilon$ and $h$, since the interfaces are moving apart and slowing each other down. We have identified these critical amplitude ratios as a function of $h$ for all combinations of $A$ and $\epsilon$. The results were obtained by iteratively searching for a solution where the growth-rates of the two interfaces were identical after $50$ non-dimensionalized time units. This is sufficient for establishing normal-mode behavior in most cases, and little variation in the results are obtained by increasing the simulation time to $75$ time units.
The results are found in Figure \ref{Fig:ArPlot01}, \ref{Fig:ArPlot05} and \ref{Fig:ArPlot09}. The wave number is constant for all plots.
 
The results show that the critical amplitude ratio varies greatly with Atwood number. The ratio varies relatively little with $h$ for small Atwood numbers. At $A=0.1$, see Figure \ref{Fig:ArPlot01}, the difference is less than 10\% between $h=0.1$ and $h=5$. The difference is larger for the higher Atwood numbers, where an approximate difference of $0.2$ between $h=0.1$ and $h=5$ is observed. Interestingly, the adjustments due to viscosity are significant for all layer thicknesses, and for the most affected cases the difference due to viscosity is almost 50 percent of the effect of layer thickness. 

The observed growth-rate coincides with that of the smallest unstable normal mode of the configuration. However, the amplitude ratio (eigenvector) of the normal mode does not, in general, coincide with the amplitude ratio of the initial condition. In Figure \ref{Fig:ArPlot01}, \ref{Fig:ArPlot05} and \ref{Fig:ArPlot09} we have thus also plotted the corresponding amplitude ratios of the normal modes. As a general trend, we observe that the difference between the two ratios is quite small. It increases with increasing viscosity and decreases with increasing Atwood number. For $\epsilon=0.001$ the ratios are indistinguishable in the plots. Furthermore, the effect is largest for small layer thickness. This indicates that the role of transients may become important for thin highly viscous layers.   

\begin{figure} 
  \ifnum\pdfstrcmp{\articletype}{preprint}=0 
    \includegraphics[width=0.75\linewidth]{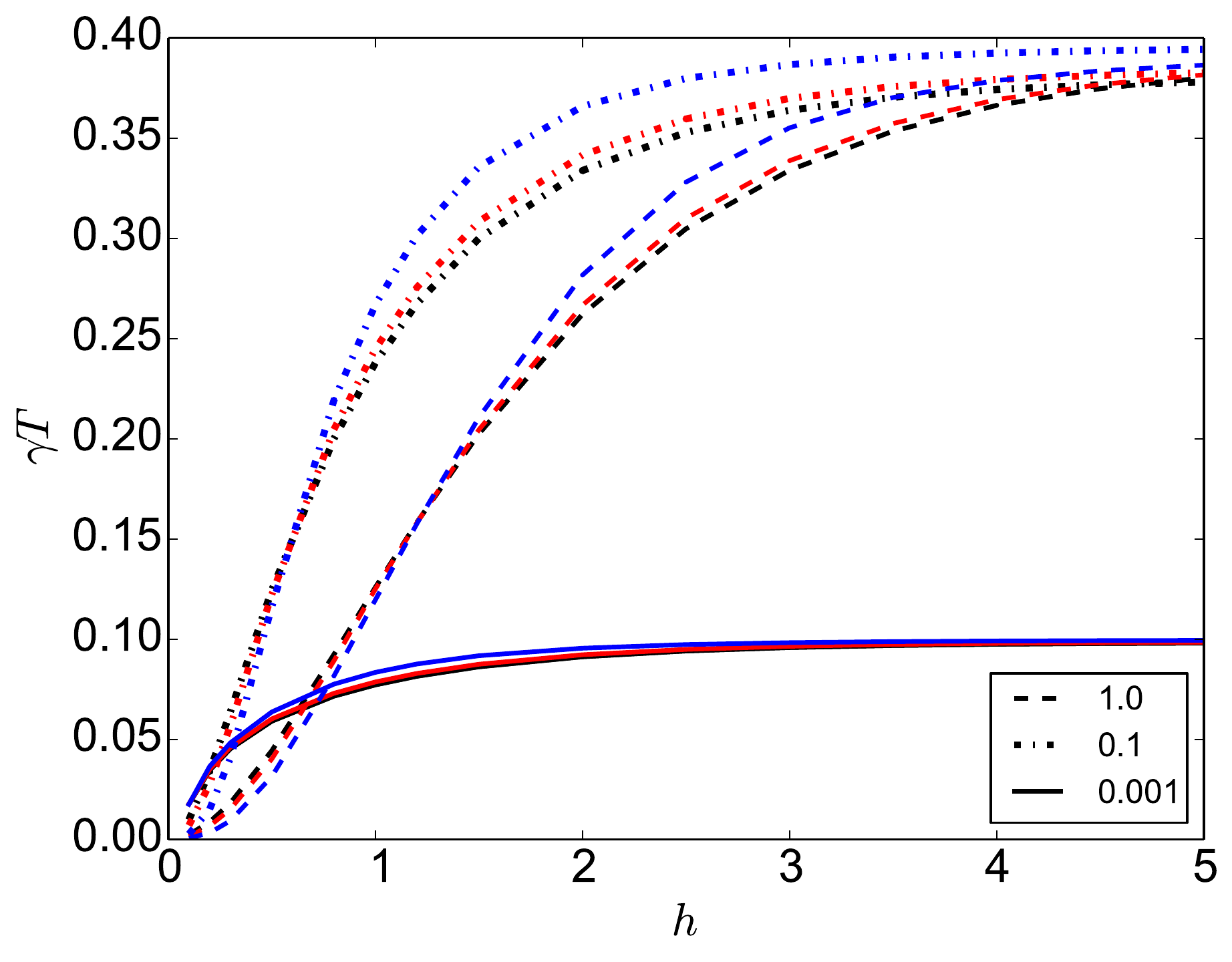}  
    \else 
    \includegraphics[width=\linewidth]{figure6.pdf}  
    \fi
 \caption{Normalized minimal growth-rate as a function of non-dimensional layer thickness ($h$) for $A=0.1$ (black), $A=0.5$ (red) and $A=0.9$ (blue). The legend denotes the value of the viscosity parameter $\epsilon$.}\label{Fig:GammaPlot09}
\end{figure}

In Figure \ref{Fig:GammaPlot09}, the growth-rates ($\gamma$), corresponding to the critical amplitude ratios, normalized by the viscous time scale ($T$), is plotted as a function of $h$. The results for the different Atwood numbers are almost identical in this scaling, with only a slight steepening of the curves for higher Atwood numbers. In contrast to the critical amplitude ratios, the dependence of the normalized growth-rate on the viscosity parameter is not monotonic, as $\epsilon=0.1$ has the largest values of the set tested here. One reason why a non-monotonic dependence on viscosity may be expected will be discussed at the end of the next section.

\subsection{Highly viscous surface layer}

An interesting limit for the above equations is what happens when a fluid layer becomes very thin. It is well known that, in the absence of surface tension the effect of such a layer becomes negligible when the layer thickness is sufficiently small \cite{Mikaelian1990b}. However, if the viscosity is very high such that $\mu H$ is appreciable one expects the effect of the layer to persist, and for sufficiently high viscosities the surface layer is expected to behave like an inextensible film \cite{Lamb1932}. 

\begin{figure} 
  \ifnum\pdfstrcmp{\articletype}{preprint}=0 
    \includegraphics[width=0.75\linewidth]{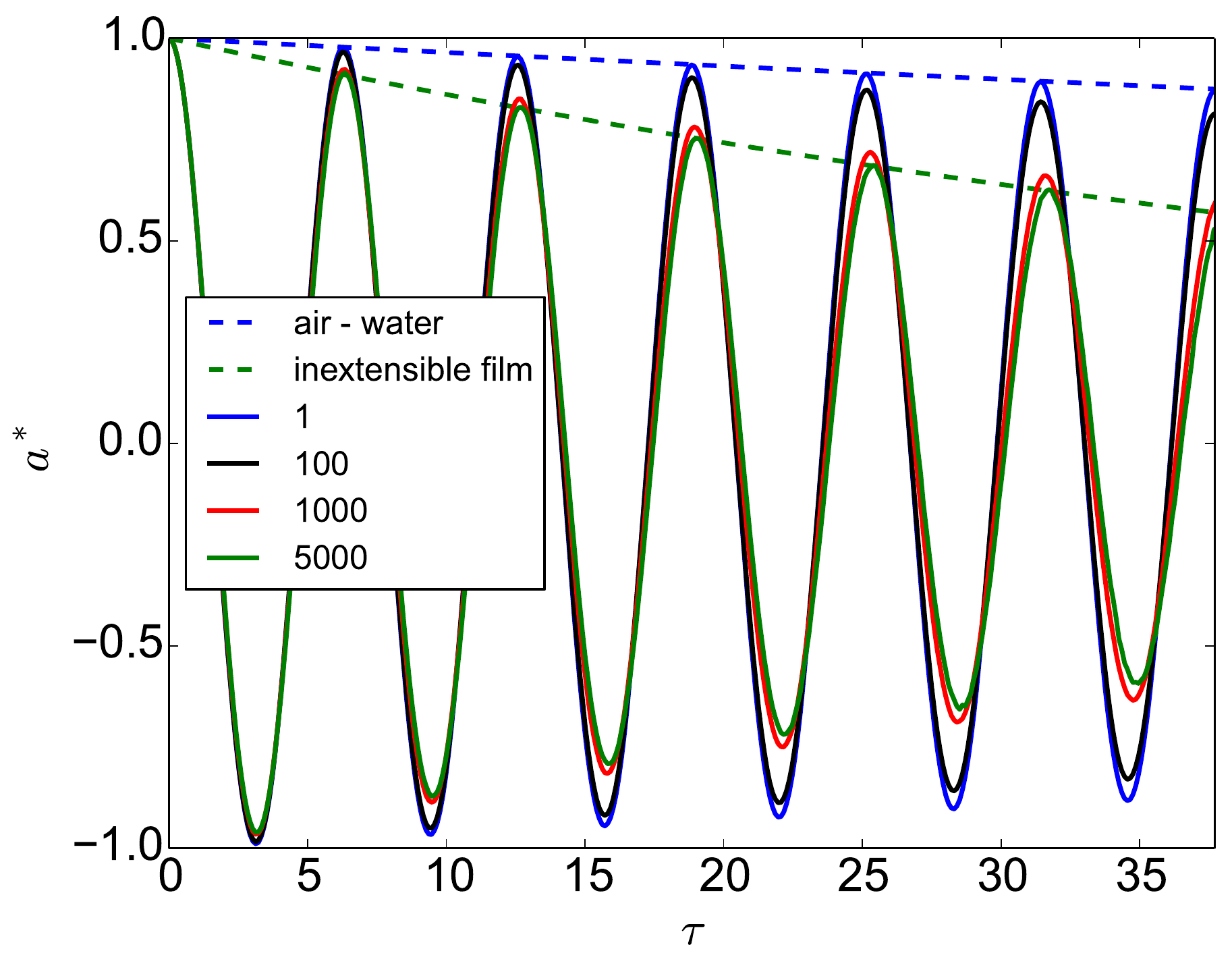}  
    \else 
    \includegraphics[width=\linewidth]{figure7.pdf}  
    \fi
 \caption{Amplitudes of the disturbances of the oil-air interface for the highly viscous surface layer case with $kH=0.01$. The legend denotes the viscosity ratio of oil to water ($\overline{\nu}=\nu_2/\nu_1$). The dashed lines are the damping rates corresponding to a pure air-water interface and an inextensible film over water.} \label{Fig:Stable01}
\end{figure}

The effect of such highly viscous surface films has also been studied in \cite{Jenkins1997}, where a dispersion relation for the stable case was derived. Here, we study a similar configuration of fluids, but for clarity the effects of surface tension is ignored. The system under consideration consists of three fluids with material properties similar to that of air, heavy oil, and water. The top (air) and bottom (water) layers have infinite extents while the middle layer has a finite thickness $H$. We consider a wave with wavelength $0.02$ m and assume that the two interfaces start with identical initial amplitudes. The various material parameters are found in Table \ref{ThinTable}. The acceleration due to gravity is set to $g=9.81$ and we non-dimensionalize time based on the inviscid natural frequency of the water-air system ($\tau=\omega t$). 
\begin{table}
\begin{center}
	\caption{Baseline fluid layer parameters for the highly viscous surface film case.} \label{ThinTable}
    \begin{tabular}{ l  l  l  l }
    \hline
    \hline
    Layer & 1 (water) \, & 2 (oil)\, & 3 (air)\, \\ \hline
    $H\, (\mathrm{m})$ & $\infty$ & $\pi^{-1} \times 10^{-4}$  & $\infty$ \\ 
    $\rho \, (\mathrm{kg/m}^3)$ & $1000$ & $900$ & $1$ \\ 
    $\nu \, (\mathrm{m}^2/\mathrm{s})$ & $10^{-6}$ & $10^{-4}$ & $10^{-5}$ \\
    \hline
    \hline
    
     \end{tabular}
\end{center}

\end{table}
As a baseline case we choose an oil layer thickness of $\pi^{-1} \times 10^{-1}$ mm, which yields $kH = 0.01$. We then vary the viscosity of the oil over several orders of magnitude. The resulting surface elevations for the oil-air interface are found in Figure \ref{Fig:Stable01}. As the viscosity is increased, the damping rate increases monotonically towards the theoretical predictions for inextensible surface films \cite{Jenkins1997}, as expected. 

\begin{figure} 
  \ifnum\pdfstrcmp{\articletype}{preprint}=0 
    \includegraphics[width=0.75\linewidth]{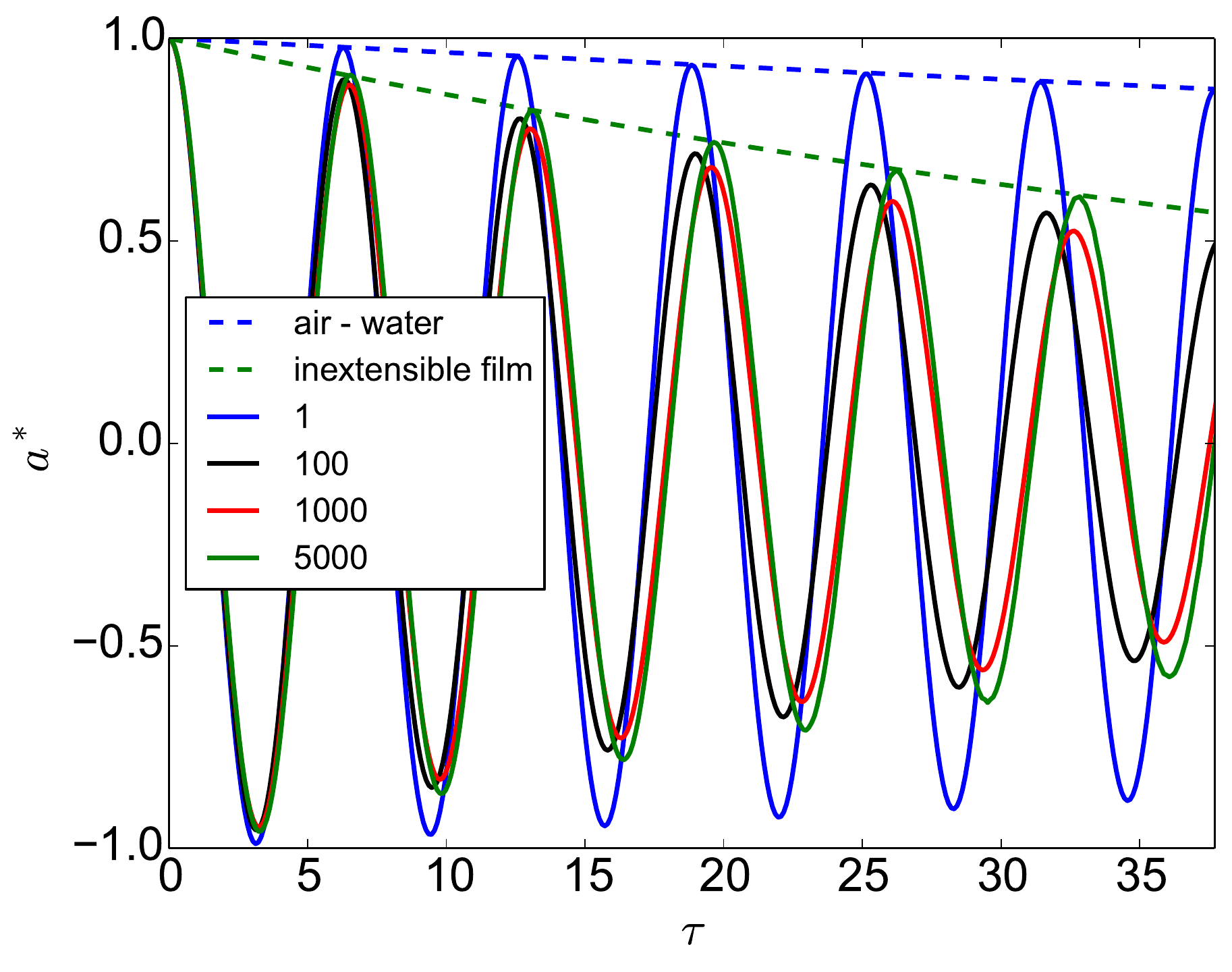}  
    \else 
    \includegraphics[width=\linewidth]{figure8.pdf}  
    \fi
 \caption{Amplitudes of the disturbances of the oil-air interface for the highly viscous surface layer case with $kH=0.1$. The legend denotes the viscosity ratio of oil to water ($\overline{\nu}=\nu_2/\nu_1$).}\label{Fig:Stable1}
\end{figure}

Next, we consider the effect of increasing the oil layer thickness. This immediately results in a loss of the monotonic increase in damping rate with increasing viscosity (seen as early as $kH=0.03$). This can be observed in Figure \ref{Fig:Stable1}, where the time histories for layer thickness $kH=0.1$ is found. In this case, the most viscous oil layers still approach the inextensible film limit, but both the curve representing viscosity ratio of $100$ and $1000$ are damped faster than this limit. This indicates that, for these viscosities, sufficient shear is generated in the middle layer to significantly contribute to the damping. As the viscosity is increased further, however, the viscous layer becomes too rigid and its damping contribution decreases.

\begin{figure} 
  \ifnum\pdfstrcmp{\articletype}{preprint}=0 
    \includegraphics[width=0.75\linewidth]{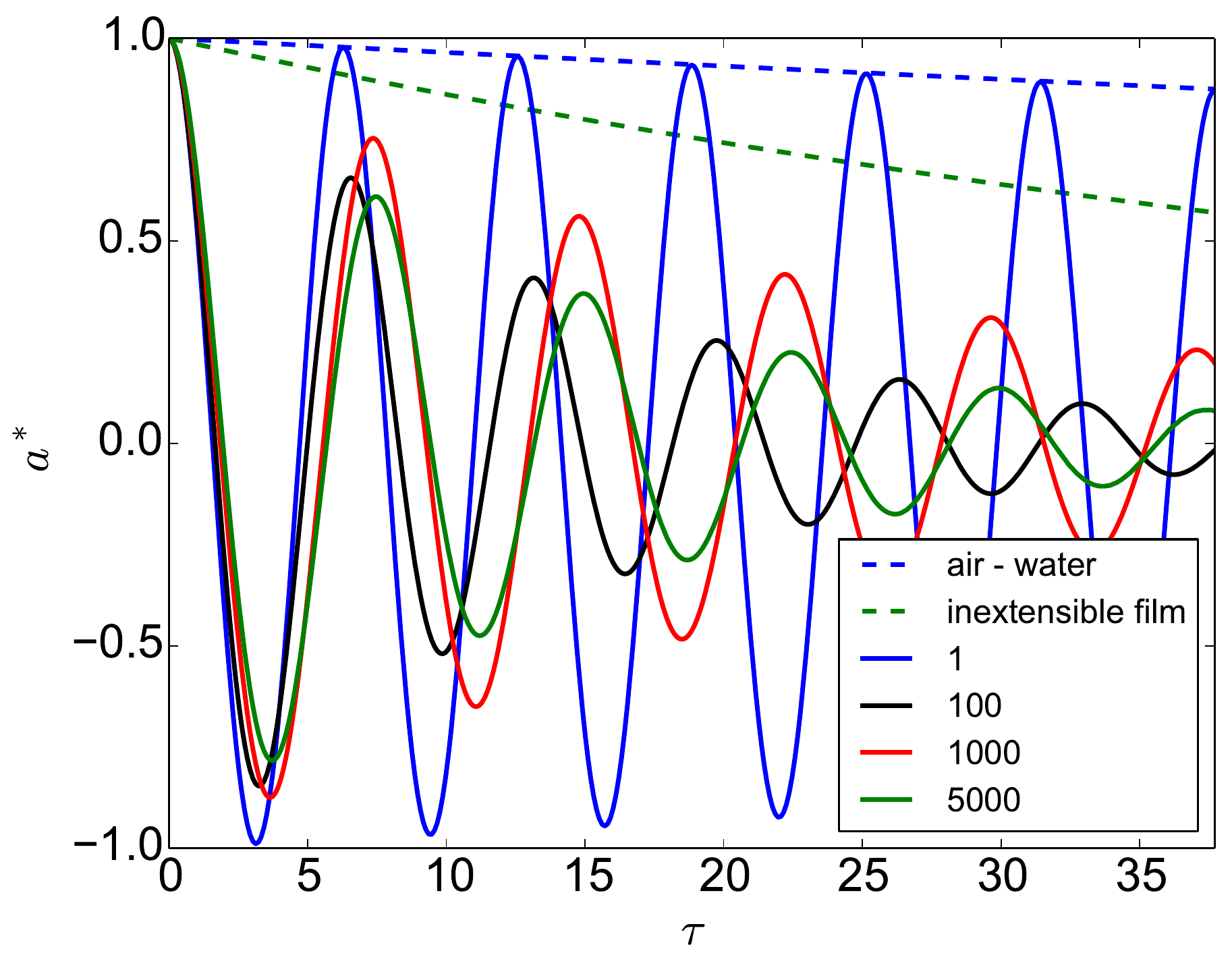}  
    \else 
    \includegraphics[width=\linewidth]{figure9.pdf}  
    \fi
 \caption{Amplitudes of the disturbances of the oil-air interface for the highly viscous surface layer case with $kH=0.6$. The legend denotes the viscosity ratio of oil to water ($\overline{\nu}=\nu_2/\nu_1$).}\label{Fig:Stable6}
\end{figure}

If we further increase the layer thickness, the results no longer converge towards the inextensible film limit when the viscosity is increased. The damping rate of the perturbation on the oil-air interface for this thicker layer is significantly higher than the damping rate for the inextensible film. In the low and high viscosity limits the damping increases with viscosity. Between these two regimes there is, however, an intermediate region where the damping decreases with increasing viscosity, and thus there exists a local minimum in the damping rate. This is seen in Figure \ref{Fig:Stable6}, which displays the time histories for $kH=0.6$, where we observe that the curve corresponding to a viscosity ratio of $1000$ has a lower damping rate than the curves corresponding to ratios of $100$ and $5000$. 

The reason for the non-monotonicity is that the vorticity diffusion terms (terms containing $\Omega$) contribute to both the stiffness and damping of the system. That this must be the case is realized by considering the behavior of the irrotational approximation for $kH=0.6$. The resulting system is over-damped ($c_i>\sqrt{m_id_i}$) above a viscosity ratio of approximately 400, and yet the interface oscillates even at a ratio of 5000. This means that the vorticity diffusion must contribute an excess stiffness great enough to alter the properties of the system. 

While the general functional dependence of the vorticity diffusion terms is complex, we can illustrate the root of the non-monotonicity by considering the following decomposition of the $\beta_i$-coefficients

\begin{equation} \label{b-decomp}
 \beta_i=\frac{2\nu_i k^2}{s} \coth(kH)+\coth(kH)-\frac{2\nu_i k \lambda_i}{s}\coth(\lambda_i H_i). 
\end{equation}
These coefficients are multiplied by the expressions for $A_i$ and $B_i$, which have the form $A_i=c_{ij}(s)\lap{\dot{a}}_j$. Here, $c_{ij}(s)$ is a matrix which depends on $s$. However, the continuity of tangential stress \eqref{lap_omega_tan} ensures that $c_{ij}$ always contains a non-zero constant component. (Note that, in the case of a viscous fluid supported between two inviscid fluids, the continuity of tangential velocity is no longer applicable, and $c_{ij}$ is a constant matrix.) 
The constant component of $c_{ij}$ contributes to the stiffness of the system when multiplied by the first term on the right hand side of \eqref{b-decomp}, while the second term acts as a pure damping term. The behavior of the third term depends on $\lambda_i$. In the limit of high viscosity $\lambda_i\rightarrow k$ and the third term cancels the added stiffness of the first. However, as $kH$ becomes small the viscosity needed to obtain this cancellation increases exponentially, as $\coth(x)$ becomes singular at 0. We thus have two competing effects where the combination of layer thickness and viscosity can cause a non-monotonic dependence on viscosity.  

As stated above, the general behavior of the vorticity diffusion terms is complex. This is a result of tangential velocity continuity \eqref{lap_omega_stress}, which introduces a non-trivial dependence of $c_{ij}(s)$ on $s$. Further investigation into this dependence is a topic for future work.
     

\section{Summary and conclusion}
In this paper, the motion of multiple superposed viscous fluids has been studied as a linearized initial-value problem. The main contribution is the development of a general closed set of equations for the Laplace transformed amplitudes of the interfaces. These equations can be inverted numerically. This formula is an extension of the single interface analysis of \cite{Prosperetti1981} to the multiple interface case. The analysis also contains the corresponding normal mode equations, which to the authors' knowledge has not been previously published. After presenting the equations we summarized the simplifications needed for including inviscid fluids, for the irrotational approximation, as well as for considering bottom and top layers of finite and infinite depth.    

The equations were used to study the effect of initial phase differences between interface perturbations on the evolution of a Rayleigh-Taylor instability and the damping effects of a highly viscous surface layer. For the Rayleigh-Taylor case we characterized the initial amplitude ratio for which the minimal possible growth-rate of the perturbations was attained as a function of viscosity, Atwood number and layer thickness. This ratio was compared to the amplitude ratio of the corresponding normal mode. The results showed that the difference in ratios increased as the fluid layers became thinner and the viscosity increased. This indicates that transients can be important for such configurations.

 For the damping of a highly viscous fluid layer case, we demonstrated that for very thin surface layers an increase in viscosity results in the system approaching the inextensible film limit.  However, as the layer thickness is increased the system quickly loses the monotonic dependence on viscosity, displaying a maximum in damping before approaching the limit. For even thicker layers the non-monotonicity persists, but the system no longer approaches the inextensible film limit.
 
Both test cases revealed that the combination of finite fluid layer thickness and a highly viscous fluid can lead to non-monotonic behavior of the interface perturbations as a function of viscosity. Analysis of the equations revealed that this non-monotonicity is a result of the vorticity diffusion contributing to both the stiffness and damping of the system. Since these contributions scale differently, both with layer thickness and viscosity, non-monotonic behavior is possible. A more comprehensive study of this non-monotonicity is a topic for future work.

\begin{acknowledgments}
The authors would like to thank Dr. Karnig O. Mikaelian for his helpful comments regarding the relationship between the normal-mode analysis and the initial-value problem. 
The authors would also like to thank Dr. Espen \AA kervik for VOF simulation results used to confirm the validity of the derived equations. 
\end{acknowledgments}

\end{document}